\documentclass[twocolumn,pre,amsmath,amssymb,showpacs]{revtex4-1}
\usepackage{graphicx,color}
\usepackage{amsmath,amssymb,latexsym,amsthm}
\usepackage{pgfplots}
\usepackage{mathtools}
\usepackage{mhchem}
\usepackage{tikz}
\usepackage{tcolorbox}
\usetikzlibrary{backgrounds}
\usetikzlibrary{chains,shapes,fit,calc,arrows}
\usepackage{pgfplotstable}

\usepackage{color}

\newcommand{\mean}[1]{{\left< #1 \right>}}

\definecolor{webgreen}{rgb}{0,.5,0}
\definecolor{webbrown}{rgb}{.6,0,0}
\definecolor{grigio}{rgb}{.85,.85,.85} 
\definecolor{RoyalBlue}{rgb}{0.0, 0.14, 0.4}
\definecolor{skyblue1}{rgb}{0.45,0.62,0.81}
\definecolor{skyblue2}{rgb}{0.2,0.39,0.64}
\definecolor{skyblue3}{rgb}{0.13,0.29,0.53}
\definecolor{scarlet1}{rgb}{0.93,0.16,0.16}
\definecolor{scarlet2}{rgb}{0.8,0,0}
\definecolor{scarlet3}{rgb}{0.64,0,0}

\usepackage[normalem]{ulem}
\usepackage{hyperref}
\hypersetup{%
    colorlinks=true, linktocpage=true, pdfstartpage=1, pdfstartview=FitV,%
    breaklinks=true, pdfpagemode=UseNone, pageanchor=true, pdfpagemode=UseOutlines,%
    plainpages=false, bookmarksnumbered, bookmarksopen=true, bookmarksopenlevel=1,%
    hypertexnames=true, pdfhighlight=/O,
    urlcolor=webbrown, linkcolor=RoyalBlue, citecolor=webgreen, 
    pdftitle={},%
    pdfauthor={},%
    pdfsubject={},%
    pdfkeywords={},%
    pdfcreator={pdfLaTeX},%
    pdfproducer={LaTeX REVTeX}%
  }

\begin{document}
\title{Local detailed balance across scales: from diffusions to jump processes and beyond}
\newcommand\unilux{\affiliation{Complex Systems and Statistical Mechanics, Physics and Materials Science Research Unit, University of Luxembourg, L-1511 Luxembourg}}
\author{Gianmaria Falasco}
\email{gianmaria.falasco@uni.lu}
\author{Massimiliano Esposito}
\email{massimiliano.esposito@uni.lu}
\unilux

\begin{abstract}
Diffusive dynamics in presence of deep energy minima and weak nongradient forces can be coarse-grained into a mesoscopic jump process over the various basins of attraction. Combining standard weak-noise results with a path integral expansion around equilibrium, we show that the emerging transition rates satisfy local detailed balance (LDB). Namely, the log ratio of the transition rates between nearby basins of attractions equals the free-energy variation appearing at equilibrium, supplemented by the work done by the nonconservative forces along the typical transition path. When the mesoscopic dynamics possesses a large-size deterministic limit, it can be further reduced to a jump process over macroscopic states satisfying LDB. The persistence of LDB under coarse-graining of weakly nonequilibrium states is a generic consequence of the fact that only dissipative effects matter close to equilibrium.
\end{abstract}


\maketitle

\section{Introduction} \label{sec:intro}

Stochastic thermodynamics is establishing itself as a comprehensive framework for the description of small systems far from equilibrium \cite{sek10, sei12, van15}. Defining thermodynamic quantities like heat, work and entropy at the level of single stochastic trajectories allows one to derive constraints on their statistics in the form of fluctuation theorems \cite{har07, esp09, jar11, rao18}, to quantify the cost of measurements and feedback \cite{hor14,par15}, and to bound  the precision \cite{bar15a, hor17, pro17, dec18, dit18, fal20a, van20} and speed \cite{shi18, ito18, nic20, ito20, fal20} of a process.
Notwithstanding its large domain of applicability, ranging from the quantum to the biochemical realm, stochastic thermodynamics is 
limited by the fact that all non-described degrees of freedom need to be equilibrated and subsumed into thermal baths.
This hypothesis is formally implemented by the condition of local detailed balance (LDB): the log ratio of the forward and backward transition rates between two states equals the entropy flow in the bath causing such transition \cite{esp12, bau14}. Crucially,  it allows to directly construct thermodynamics on top of the state dynamics, without the need of any further information.

This assumption is arguably very legitimate in those situations where the stochastic description is fundamental within the level of complexity it aims to describe. Namely, all driven degrees of freedom are explicitly described and the coarse-grained ones are singled out by a large separation in, e.g., time and length scales.
 For example, consider a bead dragged (by e.g. an optical tweezer) in a fluid described by a Langevin equation.
 The dissipation of the hydrodynamic flow field resulting from the bead motion is fully captured by the friction force and is inconsequential to the molecular degrees of freedom of the fluid \cite{fal16c}. Hence, for the driving speeds typically accessible in experiments, the fluid molecules will remain in equilibrium behaving as a thermal bath for the bead, so that LDB can be safely assumed in the Langevin equation.
 If instead the bead moves through an active \cite{cla08} or aging medium \cite{gom12}, informations about the heat dissipated by the nonequilibrium environment cannot be retained only by a carefully coarse-grained description of the bead dynamics, for which the LDB does not hold in general \cite{zam05}.
In the same spirit, the roto-vibrational states of molecules undergoing elementary chemical reactions in solution come rapidly to equilibrium with the solvent \cite{owr94} so that LDB can be used in the chemical master (resp. rate) equation for the evolution of the numbers of molecules (resp. concentrations) \cite{pri49, gil77}. Zooming out to a whole chemical network, fast (e.g. enzymatic) reactions can be adiabatically eliminated \cite{sin09} but often at the cost of misestimating their associated dissipation. Indeed coarse-graining them results in non-elementary kinetic equations (e.g. Michaelis-Menten, Hill functions) that in general do not respect LDB \cite{wac18, ava20}.

Therefore, it is important to gain basic understanding of how the LDB survives (or even emerges \cite{bau14}) under coarse graining. In this work, we first focus on coarse graining the diffusion in a multi-well potential and nonconservative force field $f$. When the temperature $T$ is low enough and $f=0$, as is well known, the equilibrium dynamics can be reduced to random independent jumps between the potential minima, whose associated rates are given by the Arrhenius-Eyring-Kramers formula \cite{han90} and satisfy LDB. For small $f$, we show in Sec.~\ref{sec:cgFP} that the coarse graining onto the mesoscopic Markov jump process remains valid and that the resulting nonequilibrium transition rates still satisfy LDB. A two-dimensional bistable system under the action of a shear force is used in Sec.~\ref{sec:ex} to illustrate the theory. In Sec.~\ref{sec:comp}, we discuss the importance of the order of the limits $f \to 0$ and  $T \to 0$ for the validity of the LDB.
We finally outline in Sec.~\ref{sec:cgME} how the LDB survives a further coarse graining onto a macroscopic jump process between a subset of states, when the mesoscopic dynamics admits a large-size limit.
The considerations in this paper hold when a global small parameter exists that defines a proper weak-noise limit. We will not discuss other types of coarse graining which are often relevant, such as adiabatic eliminations or lumping (see \cite{bo17} and references therein), whose thermodynamics has also received much attention \cite{rah07,pug10,esp12,bo14,esp15,pol17,her20}, and which may lead to non-Markovian dynamics \cite{str19}, e.g., in case of correlated recrossings between metastable states \cite{bec12}.

\section{Coarse-graining the diffusive dynamics} \label{sec:cgFP}

\subsection{Stochastic thermodynamics of diffusion}

We consider the Langevin dynamics in $\mathbb{R}^d \ni r$ 
\begin{align}\label{le}
\dot r= \underbrace{- \mu \nabla U(r) + \mu f(r)}_{\mu F(r)} +\sqrt{2 D} \xi,
\end{align}
 where $U(r)$ is a bounding potential energy, $f(r)$ is a nongradient force and $\xi$ is a zero-mean Gaussian white noise (in units of Boltzmann constant $k_\text{B}$ equal to 1).
 The stochastic dynamics \eqref{le} can be described from two other equivalent standpoints. First, by
 the Fokker-Planck equation for the probability $p(r,t)$ that the system is in $r$ at time $t$
 \begin{align}\label{fpe}
 \partial_t p(r,t)= - \nabla \cdot \underbrace{ [ \mu F(r) p(r,t) -D \nabla p(r,t) ]}_{j(r,t)},
 \end{align} 
 where $j(r,t)$ is the probability current. Second, by the conditional path probability \cite{wie86}
\begin{align}\label{path}
P[\omega|r(0)]=e^{-\frac{1}{4 D}\int_{0}^t d\tau  \{ [\dot r(\tau ) - \mu F(r(\tau ))]^2 +2 \mu D \nabla \cdot F(r(\tau)) \} }
\end{align}
for trajectories $\omega=\{r(\tau):  0 < \tau \leq t \}$ of length $t$ starting from the initial condition $r(0)$. 

A thermodynamic description can be built on \eqref{le} assuming that the mobility $\mu$ and the diffusion coefficient $D$ are connected by the Einstein relation $D= T \mu$. Here, $T$ is the temperature of the thermal bath providing both a friction force $- \dot r/\mu $ and velocity fluctuations $\sqrt{2 \mu T} \xi$. With this identification, the system enjoys LDB: the ratio of probabilities for a trajectory $\omega$ and its time-reversed one $\tilde \omega=\{r(t-\tau):  0 < \tau \leq t \}$ equals the exponential of the entropy flow into the thermal bath \cite{sei12},
\begin{align}\label{ldb}
\frac{P[\omega |r(0)]}{P[\tilde \omega |r(t)]}=e^{\frac{1}{T} \int_0^t d \tau \dot r(\tau) \cdot F(r(\tau))} =:e^{S_e[\omega]}.
\end{align}
The entropy flow $S_e= - \Delta U /T + W/T  $ is made of two contributions involving the energy difference  between the final and initial state, $\Delta U:= U(r(t))- U(r(0))$,  and the work done by the nonconservative force along the path, $W[\omega]:= \int_0^t d \tau \dot r(\tau) \cdot f(r(\tau))$.

Equation \eqref{ldb} allows to define the entropy production $\Sigma[\omega]$ by weighting the initial state of the trajectories $\omega$ (resp. $\tilde \omega$) with the probability solution of \eqref{fpe} at time 0 (resp. $t$),
\begin{align}
\frac{P[\omega] }{P[\tilde \omega]}=\frac{P[\omega |r(0)] p(r(0),0) }{P[\tilde \omega |r(t)] p(r(t),t)}=e^{S_e + \Delta S} =:e^{\Sigma[\omega]},
\end{align}
and identifying  $\Delta S := -\log p(r(t),t)+\log p(r(0),0)$ as the difference in the stochastic entropy of the system between the beginning and the end of the trajectory. For $f=0$, any initial distribution $p(r,0)$ relaxes in the long time limit to the Gibbs-Boltzmann distribution 
\begin{align}\label{peq}
p_\text{eq}(r) = \frac{1}{Z}e^{- \frac{U(r)}{T}},
\end{align} 
which is the only stationary solution of \eqref{fpe} corresponding to zero current $j_\text{eq}(r)=0$ for all $r$. This state satisfies the condition of (global) detailed balance, i.e. $\Sigma[\omega]=0$ for all trajectories $\omega$: the entropy flux in the thermal bath is exactly compensated by the variation in the system entropy,
\begin{align}\label{rev}
S_e[\omega]|_{f=0}=- \Delta U /T=\log p_\text{eq}(r(t))-\log p_\text{eq}(r(0)).
\end{align}

 \subsection{Dynamics in the weak-noise limit}\
  
To proceed with the coarse graining, we consider a potential $U(r)$ that has $N$ nondegenerate local minima $\bar r_i$, i.e. $\nabla U(\bar r_i) =0$ and $H_U(\bar r_i) > 0$, where $H_U(r):=\text{det} \left [\nabla\nabla U(r) \right]$ is the Hessian determinant of $U$ at $r$. A minimum $\bar r_i$ can be separated from nearby minima by distinct saddle points, labelled by the index $\nu$ and located at $r^{(\nu)}_{i}$. We do not impose any condition on the functional form of $f$ apart from requiring its module to be small, in a sense that will be specified later.
 
The leading order of the path probability \eqref{path} in the weak-noise limit,
\begin{align}\label{pathT0}
P[\omega|r(0)] \underset{T \to 0}{\simeq} e^{-\frac{1}{4 \mu T}\int_{0}^t d\tau  \{ [\dot r(\tau ) - \mu F(r(\tau ))]^2 }=: e^{\frac 1 T \mathcal{A}_f[\omega]},
\end{align}
can be written as the exponent of an action $\mathcal{A}_f$ whose dependence on $f$ is explicitly indicated. The most likely trajectories are found by extremizing the action in \eqref{pathT0}.
This is most easily done by switching from the present Lagrangian picture to the Hamiltonian one. To this end, we perform a Hubbard-Stratonovich transformation of \eqref{pathT0}, i.e., we introduce for all $\tau \in [0,t]$ the auxiliary (momentum) variables  $p(\tau) \in \mathbb{R}^d$ which can be removed by a functional Gaussian integration,
\begin{align}
e^{\frac 1 T \mathcal{A}_f[\omega]}&= \int \mathcal{D} p \, e^{\int_0^t d\tau \left[-\mu T p^2(\tau) -i p(\tau) \cdot (\dot r(\tau) -\mu F(r(\tau))) \right] } \nonumber \\
&=\int \mathcal{D}p\, e^{ \frac{1}{T}\int_0^t d \tau \left[-\dot r \cdot p  + H(r,p) \right] } \label{HSaction} ,
\end{align}
where $ \mathcal{D}p$ is the appropriately normalized functional measure \cite{kam11}.
In the last step we changed variable $i p T \to p$ and defined the Hamiltonian $H(r,p)= \mu p^2+ \mu p \cdot F(r)$. Extremizing the transformed action gives the equations of motion
\begin{align}\label{dr}
\dot r = \mu F + 2 \mu p \\
\dot p = -\mu \nabla F \cdot p.
\label{dp}
\end{align}
Note that solving \eqref{dr} for $p$ and plugging into \eqref{dp} we would go back to the Lagrangian picture:
\begin{align}
\ddot r = \frac{\mu^2}{2} \nabla F^2 + \mu \dot r \cdot \nabla F - \mu  \nabla F \cdot \dot r.
\end{align}
A first class of solutions of \eqref{dr} is obtained by setting $p=0$. This gives the Langevin equation \eqref{le} at $T=0$, which we call the deterministic (or noiseless, or relaxational) dynamics. To identify a second class of solutions we note that all trajectories that pass through the fixed points of $F$ are characterized by $H=0$, since $H$ is null on $\bar r_i$ and is a constant of motion. In the case $f=0$, which will be useful in the next section, the condition $H=0$ yields the solution $p= \mu \nabla U$ (in addition to $p=0$) which inserted in \eqref{dr} gives the fluctuating trajectory $\dot r= \mu \nabla U$. 


We then define the basin of attraction $B_i$ as the set of initial conditions $r(0)$ for which the deterministic dynamics  
\begin{align}\label{leeq}
\dot r^\downarrow= - \mu \nabla U(r^\downarrow) 
\end{align}
has the long time solution $\lim_{t\to \infty} r^\downarrow(t)=\bar r_i$. This time-dependent solution $r^\downarrow(t)$ of \eqref{leeq} nullifies the equilibrium action $\mathcal{A}_\text{eq}:= \mathcal{A}_f|_{f=0}$, i.e. $\mathcal{A}_{\mathrm{eq}}|_{r^\downarrow}=0$. This means that in the limit $T \to 0$ and for $f=0$ it corresponds to the most probable trajectory with initial condition $r(0)$ and final condition $\bar r_i$. The time-reversed trajectory, solution of 
\begin{align}\label{instanton}
\dot r^\uparrow= \mu \nabla U(r^\uparrow)
\end{align}
also maximizes $\mathcal{A}_\text{eq}$ (which takes the value of the Arrhenius factor $\mathcal{A}_{\mathrm{eq}}|_{r^\uparrow}=- \int d r^\uparrow \cdot  \nabla U = -\Delta U$) and corresponds to the most likely fluctuating path leading from $\bar r_i$ to $r(0)$, called instanton \cite{bra89, dyk94,tou09, bou16, bou16b}.
The definition of coarse-grained states only in terms of the potential $U$, and not of the entire force field $F$, will be \emph{a posteriori} motivated.

The resulting picture is the following: Under the assumptions that $T\to 0$ and $f\to 0$ the dynamics consists of rare excursions along $r^\uparrow$ out of the basins of attraction $B_i \subset \mathbb{R}^d$ relative to the minima $\bar r_i$, followed by fast intra-well relaxations along $r^\downarrow$ and negligibly small fluctuations around the minima. Namely, the diffusion process \eqref{le} is well approximated by a jump process with transition rates $k^{(\nu)}_{ji}$ that are the inverse mean escape time from the domain $B_i$ to $B_j$ through the saddle point at $r^{(\nu)}_{i}$ \cite{day83, fre98},
\begin{align}\label{tau}
\frac{1}{{k^{(\nu)}_{ji}}}  := \inf \{ t \geqslant 0 \,: \, r(t)=r^{(\nu)}_{i}\}.
\end{align}
Note that $k^{(\nu)}_{ji}$ is zero if $i$ and $j$ cannot be reached from the unstable manifolds of the saddle $\nu$. 
The Fokker-Planck equation \eqref{fpe} is thus coarse-grained accordingly into the master equation 
\begin{align}\label{me}
\frac{d}{dt} \varrho_i(t)= \sum_\nu \sum_{j=1}^N [ k^{(\nu)}_{ij} \varrho_j (t) - k^{(\nu)}_{ji} \varrho_i (t)],
\end{align}
for the occupation probability  $ \varrho_i(t) := \int_{B_i} dr \, p(r,t) $ of the basin $B_i $ \cite{mor95}.

\subsection{Calculation of the transition rates ratio}

To set up a proper expansion valid for weak noise and small forcing, we provisionally rescale the temperature $T = \epsilon_T T'$ and the nonconservative field $f = \epsilon_f f'(r) $, explicitly introducing the small adimensional parameters $\epsilon_T$ and $\epsilon_f$, with the quantities $T'$ and $f'(r)$ being of order $O(1)$.
At leading order in $1/\epsilon_T$, the transition rate $k^{(\nu)}_{ji}$ is given by the probability of reaching the saddle $r_i^{(\nu)}$ in infinite time starting from the attractor $\bar r_i$ \cite{fre98}, 
\begin{align}\label{kijsmallT}
k^{(\nu)}_{ji} = \lim_{t \to \infty} p(r_i^{(\nu)}, t| \bar r_i,0 ) = p_i(r)|_{r=r_i^{(\nu)}},
\end{align}
where $p(r_i^{(\nu)}, t| \bar r_i,0 )$ is the solution of \eqref{fpe} with initial condition $p(r,0)=\delta(\bar r_i)$ in the limit $T \to 0$.
Note that the end point of the transition probability in \eqref{kijsmallT} can be any position along the deterministic trajectory leading from the saddle $r_i^{(\nu)}$ to the minimum $\bar r_i$, since this additional relaxation has zero action. The long time limit in \eqref{kijsmallT} entails that the weak-noise transition probability converges to the (quasi-)stationary probability within the basin of attraction $B_i \ni r$, denoted $p_i(r)$  \cite{gra86}. To obtain $k^{(\nu)}_{ji}$, we write $p_i(r)$ as an integral over solutions of \eqref{le} starting from the local equilibrium $p_{\text{eq},i}(r(0))$ and ending in $r$ after an infinite relaxation time, 
\begin{align}\label{pi}
p_i(r) &=\lim_{t \to \infty} \int \mathcal{D} \omega \,  p_{\text{eq},i}(r(0)) P[\omega| r(0)] \delta(r(t)-r).
\end{align}
The probability $p_{\text{eq},i}$ is the local weak-noise approximation of the equilibrium distribution \eqref{peq}, 
\begin{align}\label{peqwn}
p_{\text{eq},i}(r) \simeq \sqrt{\frac{H_U(\bar r_i)}{(2 \pi T'  \epsilon_T)^d }}  e^{-\frac{1}{T' \epsilon_T} [U(r)-U(\bar r_i)]}  \quad  r \in B_i  ,
\end{align}
corresponding to a Gaussian approximation for the partition function $Z$ in $B_i $. 
We then expand the path probability $P[\omega| r(0)]$ in \eqref{pathT0} keeping only terms up to order $O(1/\epsilon_T,\epsilon_f/\epsilon_T)$,
\begin{align}\label{expansion}
 P[\omega| r(0)] \simeq P_\text{eq}[ \omega|  r(0) ]   \left[1+ \frac{\epsilon_f}{ 2 T' \epsilon_T} \int_0^t d\tau  (\dot r + \nabla U) \cdot f' \right] .
 \end{align}
Here, $P_\text{eq}[ \omega|  r ] := P [ \omega|  r ]|_{f=0}=e^{\frac 1 T \mathcal{A}_{\text{eq}}}[\omega]$ is the equilibrium path probability starting from $r$.
 
Plugging \eqref{expansion} into \eqref{pi} and using the reversibility of the equilibrium paths, as given by Eq. \eqref{rev}, we integrate over time-reversed trajectories $\tilde \omega$ (which we rename $\omega$). This results in fixing the initial position, $r(0)=r$, and reversing the sign of the velocity,
\begin{align}
p_i(r) &\simeq \lim_{t \to \infty} \int \mathcal{D} \omega \,  p_{\text{eq},i}(r(0)) \delta(r(0)-r) \nonumber \\
&\quad \quad \times  P_\text{eq}[\omega| r(0)] \left[1+ \frac{\epsilon_f}{ 2 T' \epsilon_T} \int_0^t d\tau  (-\dot r + \nabla U) \cdot f' \right] \nonumber \\
&= p_{\text{eq},i}(r)\bigg[1 - \frac{\epsilon_f}{ T' \epsilon_T} \int_0^\infty  d\tau \mean{  \dot r \cdot f' }_{\text{eq}}  \bigg] \label{normaliz} \\
&\simeq  p_{\text{eq},i}(r) e^{- \frac{\epsilon_f}{T' \epsilon_T}  \int_0^\infty  d\tau \mean{  \dot r \cdot f' }_{\text{eq}}} \label{exponent} ,
\end{align}
where $\mean{\dots}_{\text{eq}}:= \int \mathcal{D} \omega \dots  P_\text{eq}[ \omega|  r ]$.
To get rid of the product  $\nabla U \cdot f'$ and  obtain \eqref{normaliz}, we used the identity 
\begin{align}
 \int_0^t d\tau \mean{\dot r \cdot f' + \nabla U \cdot f' }_\text{eq}   =0,
\end{align}
which follows from expanding the normalization condition $ \int \mathcal{D} \omega P[ \omega| r ]=1 $ to the relevant order. The expansion in $\epsilon_f$ justifies our previous definition of basin of attraction in terms of the potential $U$ only.

To evaluate \eqref{exponent}, we use the fact that $\epsilon_T$ is small. Therefore, the average in \eqref{exponent} is dominated by a single trajectory, i.e. the relaxation trajectory $r^\downarrow(t)$, solution of \eqref{leeq}, leading from $r$ to the minimum $\bar r_i$, and whose associated action $\mathcal{A}_\text{eq}$ is zero:
\begin{align}\label{fdx}
-\int_0^\infty d\tau \mean{ \dot r \cdot f'}_\text{eq}    = - \int_{r}^{\bar r_i} f' \cdot dr^\downarrow=\int_{\bar r_i}^r f' \cdot dr^\uparrow.
\end{align}
Thanks to the reversibility of equilibrium dynamics, \eqref{fdx} can also be seen as the work performed  by the nongradient force along the instanton, i.e. the most likely fluctuating path from the minimum $\bar r_i$ to $r$. 
Therefore, retaining terms up to order $O(1/\epsilon_T, \epsilon_f/\epsilon_T)$ and using \eqref{peqwn},  \eqref{expansion} can be approximated by
\begin{align}
p_i(r)  & \simeq p_{\text{eq},i}(r)e^{\frac{1}{T} \int_{\bar r_i}^r f \cdot  dr^\uparrow }\nonumber \\
&= \sqrt{\frac{H_U(\bar r_i)}{(2 \pi T)^d }} e^{[-U(r)+U(\bar r_i)+ \int_{\bar r_i}^r f \cdot  dr^\uparrow ]/T}  \quad  r \in B_i \label{mclennan}.
\end{align}
where we reabsorbed the bookkeeping parameters $\epsilon_T$ and $\epsilon_f$.
\begin{figure}[t]
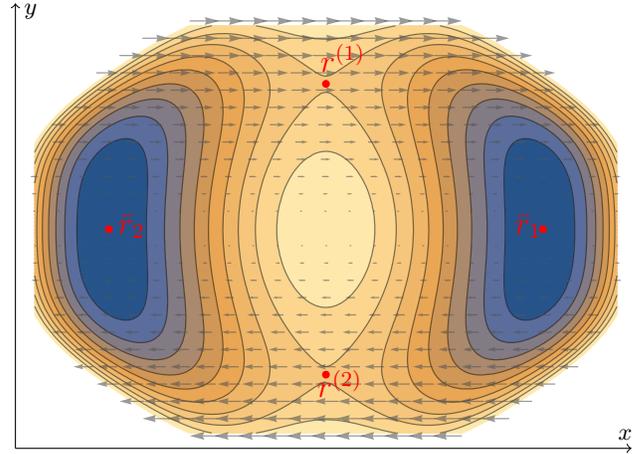


\caption{Contour plot of the potential energy \eqref{U} with the shear force field $f$ of the main text superimposed. Thanks to the parity symmetry, the local detailed balance can be verified by comparing the transition rates out of the basin of traction of only one minimum.}
	\label{fig:U}
\end{figure}

\begin{figure*}[t]
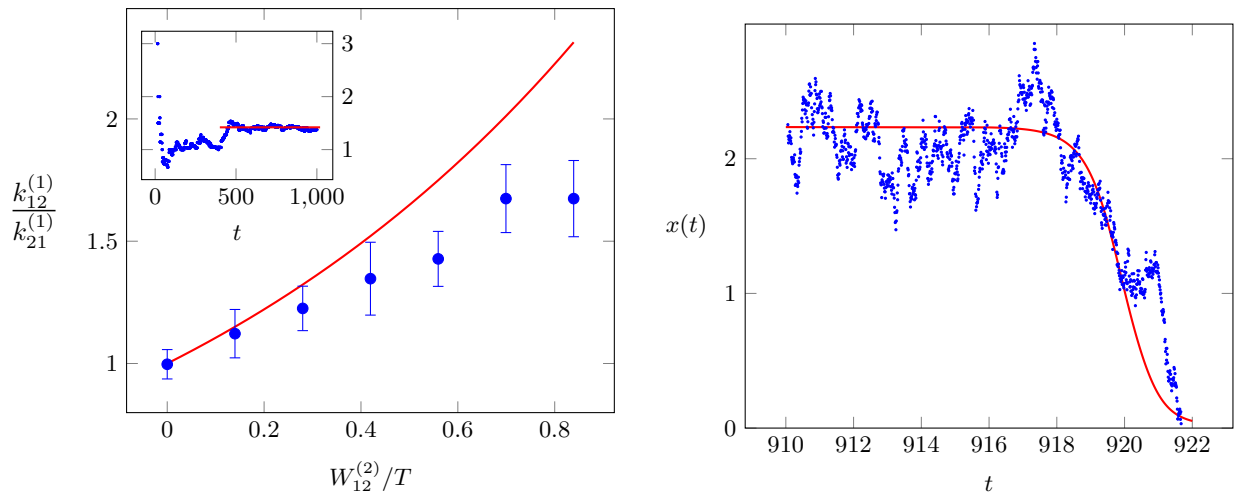

	\centering
	%
	\label{fig:ex1}
\caption{\emph{Left:} The ratio of transition rates as function of the work over temperature $W^{(2)}_{12}/T$, predicted by the theory (solid) and measured by numerical solutions of \eqref{le} obtained by the Heun scheme with $\mu =1$ and $T=0.2$. For each $\epsilon_f \in [0, 0.03]$, the time-dependent log ratio of the number of transition events across the two saddles was averaged over $10^4$ trajectories with at most $10^5$ time steps of size $\Delta t= 10^{-2}$. The procedure is repeated $n \leq 10$ times to obtain $n$ long-time averages, whose mean is taken to represent $k^{(1)}_{12}/k^{(1)}_{21}$. Error bars denote the standard deviation.
\emph{Inset:} An instance of the log ratio of the number of transition events  across the two saddles as a function of time for $\epsilon_f =0.02$. The solid line indicates the long-time average.  \emph{Right:} Projection on the $x$ axis of a typical transition path from simulations at $\epsilon_f= 0.01$ (dots) and the equilibrium instanton $x^\uparrow(t)$, i.e. the solution of \eqref{instanton} (solid), employed in the calculation of \eqref{ratioex}.}
\end{figure*}

Eventually, plugging \eqref{mclennan} into \eqref{kijsmallT}, we obtain the ratio between the transition rates involving nearby basins of attraction,
\begin{align}\label{ratio}
\frac{k_{ji}^{(\nu)}}{k_{ij}^{(\nu)}}
&= e^{[-U(\bar r_j)+U(\bar r_i) -T (S_i -S_j )  + W^{(\nu)}_{ji}]/T } .
\end{align}
In the LDB \eqref{ratio}, two thermodynamic objects appear which pertain to the coarse-grained description. First, the entropy of the coarse-grained state $i$,
\begin{align}\label{Si}
S_i:= - \frac 1 2 \log H_U (\bar r_i)
\end{align}
which is (up to an irrelevant constant shift) the Shannon entropy of the equilibrium probability \eqref{peq} under a Gaussian approximation around the minimum $\bar r_i$. Note that \eqref{Si} appears as well in the Eyring-Kramers formula for the transition rate at $f=0$  \cite{han90}, but its correspondence to the coarse-grained entropy of the state is rarely appreciated.
Second, the jump bias 
\begin{align}\label{ncw}
W^{(\nu)}_{ji}:=\int_{\bar r_i}^{r_i^{(\nu)}} f(r^\uparrow) \cdot dr^\uparrow + \int_{r_i^{(\nu)}}^{\bar r_j} f(r^\downarrow) \cdot dr^\downarrow,
\end{align}
which is the work done by $f$ along the most probable path from $\bar r_i$ to $\bar r_j$ through $r_i^{(\nu)}$, i.e. the instanton to the saddle point followed by the relaxational dynamics. Equation \eqref{ratio} is the coarse-grained analog of \eqref{ldb}, with the only difference that the energy of the diffusive dynamics is replaced by the free energy of the discrete states,
\begin{align}\label{freeenergy}
\mathcal{F}_i := U(\bar r_i)-T S_i.
\end{align}

\section{Illustrative example}\label{sec:ex}

To exemplify our result, we consider the dynamics \eqref{le} in $\mathbb{R}^2 \ni r=(x,y)$ with the double-well potential 
\begin{align}\label{U}
U(x,y) =0.1(x^2 y^2 - 10 x^2 + x^4 + y^4 - 4.5 y^2 + 0.1 x^4 y^4)
\end{align}
and the nonconservative force $f(x,y)=(\epsilon_f y, 0)$ which represents a shear of intensity $\epsilon_f$. The symmetric potential has $N=2$ minima, $\bar r_1 \simeq(2.24 ,0)$ and $\bar r_2 \simeq(-2.24 ,0)$, connected by two saddle points $r^{(1)}=(1.5,0)$ and $r^{(2)}=(-1.5,0)$ (see Fig.~\ref{fig:U}). Since \eqref{le} is invariant under a parity transformation $r \mapsto -r$, the LDB \eqref{ratio} can be evaluated focusing only on the transition rates out of one minimum, say $\bar r_1$, through the saddle points $r^{(1)}$ and $r^{(2)}$, namely,
\begin{align}\label{ratioex}
\frac{k^{(1)}_{12}}{k^{(1)}_{21}}=\frac{k^{(2)}_{21}}{k^{(1)}_{21}}= e^{W^{(2)}_{21}/T}= e^{2 \frac{\epsilon_f}{T} \int_{0}^\infty y^\uparrow(t) \dot x^\uparrow (t) dt}.
\end{align}
Here, $x^\uparrow(t)$ and $y^\uparrow(t)$ are the components of the instanton starting from  $\bar r_1$ and ending in $r^{(2)}$, which equal by symmetry the relaxation path from $r^{(2)}$ to $\bar r_2$. 
The formula \eqref{ratioex} is expected to hold for $W^{(2)}_{21} \ll T \ll U(r^{(2)})-U(\bar r_1)$. However, the comparison with the results of numerical integrations of \eqref{le} shows the qualitative agreement with the theory even at moderate values of noise strength and shear work, $[U(r^{(2)})-U(\bar r_1) ]/T \simeq 10 $ and $W^{(2)}_{21}/T \lesssim 0.4$, respectively. This example indicates that the assumptions we employed in our derivations are only sufficient but may not be necessary at all in many specific cases.

\section{Discussion of the weak forcing condition}\label{sec:comp}

Our derivation hinges on the formal conditions of weak noise and weak forcing in the form $\epsilon_f \ll 1$, $\epsilon_T\ll 1$ and $\epsilon_f/\epsilon_T \ll 1$. Physically, as shown in the previous example, these requirements correspond to a small thermal energy with respect to the energy barrier, i.e. $T \ll U(r^{(\nu)}_i)-U(\bar r_i)$, and a comparatively smaller nonconservative work, i.e. $W^{(\nu)}_{ji} \ll T$. The order in which we apply these conditions, i.e. first expanding in $\epsilon_f $ and only later in $\epsilon_T$, is crucial.
 This can be inspected by comparing with  \cite{bou16}, where the local weak noise stationary distribution 
\begin{align}\label{kbou}
p_i(r) = \sqrt{\frac{H_{\phi_i}(\bar r_i)}{(2 \pi T)^d }}  e^{-\frac 1 T [\phi_i(r)-\phi_i(\bar r_i)] + \int_0^\infty dt \mu \nabla \cdot \ell(r_f^\uparrow(t))}  
\end{align}
valid for $r \in B_i $, was derived taking into account sub-exponential corrections and arbitrarily large $f$. In \eqref{kbou}, $\phi_i$ is the so called quasi-potential obtained by evaluating the action $\mathcal{A}_f$ in \eqref{path} on the forced instanton $r_f^\uparrow(t)$, i.e. the path starting in $\bar r_i$ and ending in $r \in B_i$ which maximizes $\mathcal{A}_f$; $\ell := (\nabla \phi_i + F) $ is the drift field tangent to the level sets of $\phi_i$; $H_{\phi_i}(\bar r_i)$ is the Hessian determinant of $\phi_i$ in $\bar r_i$.

In our approach, the first expansion \eqref{exponent} in $\epsilon_f \ll 1$ allows us to discard any nonequilibrium contribution to the instanton. This  is equivalent, when taking the second expansion in $\epsilon_T \ll 1$, to replace $r^\uparrow_f(t)$ with $r^\uparrow(t)$ in the calculation of $\phi_i$ and $\ell(r^\uparrow_f(t))$ in \eqref{kbou}. This yields the quasi-potential
\begin{align}
\phi_i(r) = U(r) - \epsilon_f \int_{\bar r_i}^r f' \cdot  dr^\uparrow ,
\end{align}
which implies $H_{\phi_i} = H_U + O(\epsilon_f)$, $\ell = O(\epsilon_f)$, and so our equation \eqref{mclennan}.

Note that if we exchanged the order of the limits or we included higher order terms in $\epsilon_f$, we would not be able to obtain the LDB for the log ratio of the transition rates without additional assumptions. On one hand, $\phi_i$ and $\log H_{\phi_i}(\bar r_i)$ would still represent a local potential for the dynamics \cite{gra84} and a Gaussian approximation of the Shannon entropy of state $i$, respectively. But, on the other hand, $\nabla \cdot \ell$ would not have any straightforward thermodynamic interpretation. In fact, $\ell$ can be written as either the (local) orthogonal decomposition of $F=-\nabla \phi_i + \ell$, i.e. $\nabla \phi_i \cdot \ell =0$, or the leading order of the stationary velocity in probability space, i.e. $\ell(r)=\lim_{\epsilon_T \to 0}  j(r)/p(r)$ \cite{bou16, zho16}. 
It remains to be seen whether these dynamic and probabilistic viewpoints entail any corresponding thermodynamic notion.

We conclude noting that a straightforward extension exists to the case of potentials $U(r,\lambda(t))$ that depend on time through a prescribed protocol $\lambda(t)$ \cite{tal04, luc19}. The variations of $U$ in time should have frequencies much smaller than the largest equilibration time within basins,  preserve the number $N$ of minima and be consistent with the assumptions of weak noise, i.e. $U( r_i^{(\nu)},\lambda(t))-U(\bar r_i,\lambda(t)) \gg T$ for all $i$ and $t$.  Under this condition, the escape events take place in a fixed force field, and the variations of the potential happen quasi-statically while the system fluctuates in a minimum $\bar r_i$. Thus, our derivation can be replicated as is and the LDB \eqref{ratio} acquires a parametric dependence on time through $U(r,\lambda(t))$.

\section{Towards a thermodynamic path through scales}\label{sec:cgME}

\begin{figure}[t]
	\centering 
	\includegraphics[width=0.5\textwidth]{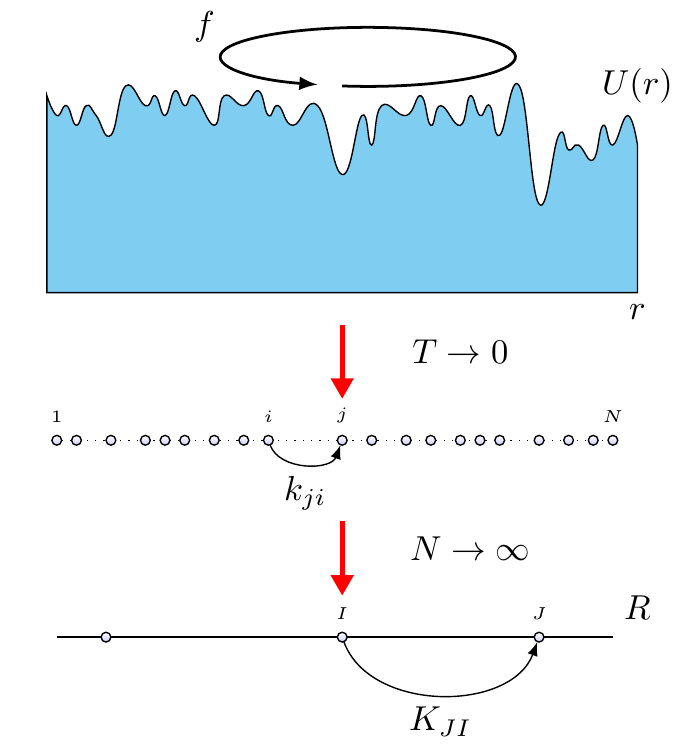}
	\caption{Schematic representation of consecutive coarse-graining levels. Diffusion in a rouged energy landscape superimposed to a small nonconservative force $f$ behaves at low temperatures as a Markov jump process with transitions rates $k_{ji}$ between the nearby basins of attraction of the energy minima. For a large number of attractors $N$ the state space becomes again continuous. Nevertheless, if transition rates scale as $N$ another weak-noise limit exists that singles out a further reduced set of states with Markovian transition rates $K_{JI}$.}
	\label{fig:sketch}
\end{figure}

The approach outlined in this work is suitable to be replicated whenever \eqref{me} can be further coarse-grained onto a more reduced set of states. This often happens when an additional large parameter exists that induces a new strong separation of time scales through a new weak-noise limit (see Fig.~\ref{fig:sketch}). Consider the case of a large state space, $N \to \infty$, so that a continuous variable $R=i/N$ can be introduced. If the microscopic energy $U$ is such that the non-zero transition rates behave asymptotically as $k^{(\nu)}_{ji} \sim  k^{(\nu)}(R)$ with $k^{(\nu)}(R)$ of order $O(N)$, the stationary probability scales as $N \varrho_i = \rho(R) \sim e^{-N \psi(R)}$ and thus concentrates on the minima $I= 1, \dots, M$ of the quasi-potential $\psi(R)$ \cite{gan87}, which at zero forcing equals the free energy `density' $\mathfrak{F}(R)=\lim_{N \to \infty} \mathcal{F}_i/N$. The long time dynamics is a Markov jump process between such minima with transition rates $K_{JI}$ estimated in analogy with \eqref{kijsmallT} as
\begin{align}\label{KIJsmallT}
K_{JI}  
= \rho_I(R)|_{R=R_I^{(\nu)}}.
\end{align}
Here $\rho_I(R)$ is the local stationary solution of the (weak-noise) continuous-space limit of \eqref{me}, i.e. the solution of
\begin{align}
0&= \sum_\nu \left[ k^{(\nu)}\left(R-\frac{\varepsilon^{(\nu)}}{N} \right)\rho \left (R-\frac{\varepsilon^{(\nu)}}{N} \right)- k^{(\nu)}(R)\rho(R)\right] \nonumber \\
&  \underset{N \to \infty}{\simeq} \sum_\nu \left[ (e^{\varepsilon^{(\nu)} \cdot \partial_R \psi(R)} -1) k^{(\nu)}(R)  \right] \label{merho}
\end{align}
where $\varepsilon^{(\nu)}$ equals the distance between $i$ and $j$ if connected by the saddle $\nu$, and 0 otherwise. Equation \eqref{merho} is a time-independent Hamilton-Jacobi equation for the position $R$ and momentum $\partial_R \psi(R)$ \cite{gan87, dyk94}. This is very analogous to what discussed in Sec. \ref{sec:cgFP} for low-temperature diffusion.
Note that \eqref{merho} cannot be consistently expanded in a power series in $\varepsilon^{(\nu)}$ and thus truncated as a Fokker-Planck equation unless $\varepsilon^{(\nu)} $ is infinitesimal and $\sum_\nu \varepsilon^{(\nu)} k^{(\nu)}$ and $\sum_\nu {\varepsilon^{(\nu)}}^2 k^{(\nu)}$ are of the same order \cite{gardiner}.

As already done for the diffusive dynamics, \eqref{KIJsmallT} can be obtained by expanding around equilibrium (i.e. $W_{ji}^{(\nu)}=0$ for all $i, j$ and $\nu$) the path integral for the trajectory $\omega=\{(i(\tau),\nu(\tau)): 0 < \tau \leq t \}$,
\begin{align}\label{pathme}
P[\omega|i(0)]= \left( \prod_{\alpha} k^{\nu(t_\alpha)}_{i(t_{\alpha}) \, i(t_{\alpha}^-)} \right) e^{\int_0^t d \tau \sum_{j=1}^N k_{j i(\tau)}}
\end{align}
where $t_\alpha$ labels the transition times \cite{sun06}, and by taking the leading order in $N$. Repeating the very same steps of Eqs.~\eqref{pi}--\eqref{mclennan} with \eqref{path} replaced by \eqref{pathme} leads to the LDB
\begin{align}\label{ldbK}
\frac{K_{JI}}{K_{IJ}}= \frac{K_{JI}}{K_{IJ}}\bigg|_{W_{ij}^{(\nu)}=0} e^{\mathcal{W}_{JI}/T}
\end{align}
where $\mathcal{W}_{JI}= \lim_{N \to \infty} \mean{\sum_\alpha  W^{\nu(t_\alpha)}_{i(t_{\alpha}) \, i(t_{\alpha}^-)} \delta(R(0)-I)}$. The nonequilibrium correction in \eqref{ldbK} is the (mean) work along the most probable trajectory connecting the macrostate $I$ to $J$, which are minima of the free energy $\mathfrak{F}(R)$. Such path can be more easily found by a path integral representation of the probability \eqref{pathme} \cite{dyk94}. 
It is formally analogous to the nonequilibrium correction in \eqref{ratio} since in both cases a first order expansion around equilibrium for the (local) stationary probability of states was performed,  which is universally determined by the dissipative part of the dynamics \cite{col11, mae10b}. This suggests that LDB should persist whenever nonconservative forces are small on the scale we wish to apply the coarse-graining.

\section{Conclusions}

In this paper we have showcased a general method to coarse grain a diffusive dynamics with weak noise and small nonconservative forces into a jump process that satisfies local detailed balance (LDB). 
The method extends to master equations corresponding to determinist dynamics in the zero-noise limit, thanks to the universality of the first order expansion around equilibrium.
The important next step is to generalize this approach to finite forcing $f$ (resp. $W^{(\nu)}_{ji}$) where reduced states $i$ (resp. $I$) are genuine nonequilibrium ones, which requires continuous dissipation to be sustained. Namely, they emerge as local minima of the quasi-potential $\phi$ (resp. $\psi$) and cannot be anticipated by the sole knowledge of an underlying energy $U(r)$ (resp. free energy $\mathcal{F}_i$). In order to describe the thermodynamics of transitions between them we expect to give up the LDB unless special conditions are met, i.e. to renounce the idea that thermodynamics is fully determined by the state dynamics, and to derive coarse-grained dynamical equations for the entropy production and other thermodynamic observables.

\bibliography{draftLDB}

\end{document}